\documentclass{emulateapj-rtx4}

\slugcomment{}

\shorttitle{Successive Merger of Massive Black Holes}
\shortauthors{Tanikawa and Umemura}

\begin{document}

\title{Successive Merger of Multiple Massive Black Holes in a
  Primordial Galaxy}

\author{A. Tanikawa\altaffilmark{1} and M. Umemura\altaffilmark{1}}
\affil{$^1$Center for Computational Sciences, University of Tsukuba,
  1-1-1, Ten-nodai, Tsukuba, Ibaraki 305-8577, Japan}

\begin{abstract}
Using highly-accurate $N$-body simulations, we explore the evolution
of multiple massive black holes (hereafter, MBHs) in a primordial
galaxy that is composed of stars and MBHs. The evolution is pursed
with a fourth-order Hermite scheme, where not only three-body
interaction of MBHs but also dynamical friction by stars are
incorporated. Initially, ten MBHs with equal mass of $10^7M_\odot$ are
set in a host galaxy with $10^{11}M_\odot$. It is found that $4$ - $6$
MBHs merge successively within 1 Gyr, emitting gravitational wave
radiation. The key process for the successive merger of MBHs is the
dynamical friction by field stars, which enhances three-body
interactions of MBHs when they enter the central regions of the
galaxy.  The heaviest MBH always composes a close binary at the
galactic center, which shrinks owing to the angular momentum transfer
by the third MBH and eventually merges. The angular momentum transfer
by the third MBH is due to the sling-shot mechanism.  We find that the
secular Kozai mechanism does not work for a binary to merge if we
include the relativistic pericenter shift.  The simulations show that
a multiple MBH system can produce a heavier MBH at the galactic center
purely through $N$-body process.  This merger path can be of great
significance for the growth of MBHs in a primordial galaxy.  The
merger of multiple MBHs may be a potential source of gravitational
waves for the Laser Interferometer Space Antenna (LISA) and pulsar
timing.
\end{abstract}


\section{Introduction}

In the central regions of galaxies, massive black holes with $10^6$ -
$10^9 M_{\odot}$ (hereafter, MBHs) are found to reside. They are
believed to coevolve with their host galaxies, since the masses of
MBHs correlate with the mass and velocity dispersion of the spheroidal
components of the host galaxies
\citep{Kormendy95,Magorrian98,Ferrarese00,Tremaine02,Marconi03}.  Such
MBHs seem to acquire most of their masses through gas accretion
process in the final evolutionary stage \citep{Soltan82,Yu02}. But,
the growth from an order of magnitude smaller black holes is still
unresolved.

In the last decade, many quasars at redshift $z \sim 6$ have been
observed \citep[e.g.,][]{Fan01}. This suggests that MBHs with $10^9
M_{\odot}$ have formed when the age of the Universe is only $1$
Gyr. If seed black holes are the remnants of first stars with $\sim
100M_{\odot}$ \citep{Abel00,Nakamura01,Bromm02,Yoshida06}, the seed
black holes cannot grow to $10^9M_{\odot}$ with continuous Eddington
accretion rate over $1$ Gyr. One solution is the growth by
super-Eddington accretion \citep[e.g.,][]{Abramowicz88,Ohsuga05}.
However, gas accretion onto the seeds should be intermittent, and on
average could be lower than the Eddington accretion
\citep{Milosavljevic09a,Milosavljevic09b}.

Hence, it is worth considering the possibility of black hole mergers
for the growth of MBHs.  In the cold dark matter cosmology, a massive
galaxy forms through the multiple merger of subgalaxies.  If
subgalaxies possess MBHs, a massive galaxy should contain multiple
MBHs shortly after the merger.  On the other hand, besides some
candidates of binary MBHs \citep[e.g.,][]{Sudou03,Boroson09,Dotti09},
there is few evidence for multiple MBHs in a massive galaxy.  Thus,
multiple MBHs possibly merge into a heavier BH.  But, the merger path
of multiple MBHs in a massive galaxy has not been hitherto resolved.

In this paper, we explore the evolution of the system of multiple MBHs
in a massive galaxy.  Previous studies found that if two MBHs are in
one galaxy, they are hard to merge within a Hubble time due to loss
cone depletion \citep{Begelman80, Makino04}.  If three MBHs are in one
galaxy, two of them merge or result in a binary, and the other is
ejected from the galaxy \citep{Iwasawa06}.  Here, we consider a
two-component system that consists of MBHs and stars, where not only
three-body interaction of MBHs but also dynamical friction by stars
are incorporated.  The paper is organized as follows. In section
\ref{sec:mthd}, we describe our simulation method to follow the
evolution of the multiple MBHs. In section \ref{sec:rslt}, simulation
results are presented.  In section \ref{sec:dis}, we discuss the
validity of our model. In section \ref{sec:sum}, we summarize this
paper.

\section{Method}
\label{sec:mthd}

\subsection{Setup}

We treat MBHs and stars as a $N$-body system. 
An individual MBH corresponds to one massive particle, and field
stars composing a host galaxy are approximated as super particles,
a fraction of which may be interpreted as dark matter.

The field stars are distributed according to the Hernquist model
\citep{Hernquist90}, whose radial mass density distribution,
$\rho(r)$, is given by
\begin{equation}
  \rho(r) = \frac{M}{6\pi r_{\rm v}^3} \frac{1}{(r/r_{\rm v})
    \left[(r/r_{\rm v})+1/3\right]^3},
\end{equation}
where $M$ and $r_{\rm v}$ are the total mass and virial radius of the
galaxy, respectively. The virial radius is given by
$
  r_{\rm v} = GM/4|E|,
$
where $G$ is the gravitational constant, and $E$ is the total energy
of the galaxy. The number of the field stars is $N=512K
(1K=1024)$. 

Here, ten MBHs with equal mass of $10^{7}M_{\odot}$ are set in a
galaxy with $10^{11}M_{\odot}$.  This means that the total mass of the
MBHs is $0.1$ \% of the galaxy mass and the mass ratio of each MBH to
each field star is about $50$.  We perform five simulations with
different phase space distributions of the MBHs at the initial time in
order to see the dependence on the stellar mass density around the
MBHs.  The MBHs are distributed initially within one-third, two-third,
and one virial radius of the galaxy in model A, B, and C,
respectively.  Furthermore, in model A, the positions of MBHs are
changed according to three different set of random number, which are
labeled by model A$_1$, A$_2$, and A$_3$, respectively. These models
are summarized in Table \ref{tab:sum}.

In the present simulations, we adopt the standard $N$-body units,
$G=M=r_{\rm v}=1$.  In such units, the time unit of simulation,
$t_{\rm nu}$, is comparable to the dynamical time, $ t_{\rm nu} \sim
t_{\rm dy}.  $ The light speed is $c=600$ in this units, which means
that the three-dimensional velocity dispersion of field stars is $300$
km/s at the galaxy center.

If we convert the $N$-body units to physical units, the virial radius,
the dynamical timescale within the virial radius, $t_{\rm dy}$, and
the average mass density of the galaxy within the virial radius,
$\rho_{\rm v}$, are respectively given by
\begin{eqnarray}
  r_{\rm v}   &\sim & 2 \left(\frac{M}{10^{11}M_{\odot}}\right)
  \mbox{[kpc]}, 
\label{scaling1}\\
  t_{\rm dy}  &\sim & 6 \left(\frac{M}{10^{11}M_{\odot}}\right)
  \mbox{[Myr]}, 
\label{scaling2}\\
  \rho_{\rm v} &\sim & 1 \left(\frac{M}{10^{11}M_{\odot}}\right)^{-3}
  \mbox{[$M_{\odot}\mbox{pc}^{-3}$]}.
\label{scaling3}
\end{eqnarray}
According to this scaling, the present results can be applicable for
the different mass system, which is discussed later.

\subsection{Merger condition}

We assume that two MBHs merge through gravitational wave radiation,
when the separation between two MBHs is less than ten times the sum of
their Schwarzschild radii:
\begin{equation}
  \left|{\bf r}_{{\rm B},i} - {\bf r}_{{\rm B},j} \right| < 10 \left(
  r_{{\rm sch},i} + r_{{\rm sch},j} \right),
\end{equation}
where ${\bf r}_{{\rm B},i}$ and $r_{{\rm sch},i}$ are the position and
Schwarzschild radius of $i$-th MBH. The Schwarzschild radius of $i$-th
MBH is $r_{{\rm sch},i} = 2Gm_{{\rm B},i}/c^2$ for the MBH mass
$m_{{\rm B},i}$.

\subsection{Equation of motion}

The equations of motion for field stars and MBHs are respectively
given by
\begin{eqnarray}
  \frac{d^2{\bf r}_{{\rm f},i}}{dt^2} &=& \sum_{j \neq i}^{N_{\rm f}}
       {\bf a}_{{\rm ff},ij} + \sum_{j}^{N_{\rm B}} {\bf a}_{{\rm
           fB},ij} \\ \frac{d^2{\bf r}_{{\rm B},i}}{dt^2} &=&
       \sum_{j}^{N_{\rm f}} {\bf a}_{{\rm Bf},ij} + \sum_{j \neq
         i}^{N_{\rm B}} {\bf a}_{{\rm BB},ij},
\end{eqnarray}
where ${\bf r}_{{\rm f},i}$ and ${\bf r}_{{\rm B},i}$ are the
positions of $i$-th field star and MBH, $N_{\rm f}$ and $N_{\rm B}$
are the numbers of field stars and MBHs, ${\bf a}_{{\rm ff},ij}$ and
${\bf a}_{{\rm fB},ij}$ are the accelerations of $j$-th field star and
MBH on $i$-th field star, and ${\bf a}_{{\rm Bf},ij}$ and ${\bf
  a}_{{\rm BB},ij}$ are the accelerations of $j$-th field star and MBH
on $i$-th MBH, respectively.  Excepting the MBH-MBH interaction, the
accelerations are given by Newtonian gravity:
\begin{eqnarray}
  {\bf a}_{{\rm ff},ij} &=& - Gm_{{\rm f},j}\frac{{\bf r}_{{\rm f},i}
    - {\bf r}_{{\rm f},j}}{(|{\bf r}_{{\rm f},i} - {\bf r}_{{\rm
        f},j}|^2+\epsilon^2)^{3/2}} \\ {\bf a}_{{\rm fB},ij} &=& -
  Gm_{{\rm B},j}\frac{{\bf r}_{{\rm f},i} - {\bf r}_{{\rm B},j}}{|{\bf
        r}_{{\rm f},i} - {\bf r}_{{\rm B},j|^3}} \\ {\bf a}_{{\rm
          Bf},ij} &=& - Gm_{{\rm f},j}\frac{{\bf r}_{{\rm B},i} - {\bf
          r}_{{\rm f},j}}{|{\bf r}_{{\rm B},i} - {\bf r}_{{\rm
            f},j}|^3},
\end{eqnarray}
where $m_{{\rm f},j}$ and $m_{{\rm B},j}$ are respectively the masses
of $j$-th field star and MBH, and the softening parameter
($\epsilon=10^{-3}$) is introduced only in star-star interactions.

The acceleration between two MBHs contains Newtonian gravity and
post-Newtonian corrections, such as
\begin{equation}
  {\bf a}_{{\rm BB},ij} = - Gm_{{\rm B},j}\frac{{\bf r}_{{\rm
        B},i}-{\bf r}_{{\rm B},j}}{|{\bf r}_{{\rm B},i}-{\bf r}_{{\rm
        B},j}|^3} + {\bf a}_{{\rm PN},ij}. \label{aBB}
\end{equation}
For the second term (${\bf a}_{{\rm PN},ij}$), the pericenter shift
(1PN and 2PN terms) as well as the gravitational radiation emission
(2.5PN term) is considered \citep{Damour81,Soffel89,Kupi06}. We adopt
equation (1), (2), (3), and (4) in \cite{Kupi06} for the second term.

If the semi-major axis is less than $a_{\rm crit}=5 \times 10^{-5}$,
the motion of the binary is transformed to the motion of the center of
mass and the relative motion. We ignore tidal forces by distant field
stars on the binary. Then, the acceleration by a distant field star
$k$ to the center of mass ($a_{{\rm cm},k}$) and the relative motion
($a_{{\rm rel},k}$) is approximated as
\begin{eqnarray}
  {\bf a}_{{\rm cm},k} &\approx&
  -Gm_{{\rm f},k}\frac{{\bf r}_{\rm cm} - {\bf r}_{{\rm f},k}}{|{\bf 
      r}_{\rm cm} - {\bf r}_{{\rm f},k}|^3}, \\
  {\bf a}_{{\rm rel},k} &\approx& 0, 
\end{eqnarray}
where ${\bf r}_{\rm cm}$ is the position of the center of mass of the
MBH binary. Distant field stars are defined as
\begin{equation}
  \left| {\bf r}_{\rm cm} - {\bf r}_{{\rm f},k} \right| > Ca_{\rm crit},
\end{equation}
where $C=200$.

\subsection{Numerical scheme}

We use a fourth-order Hermite scheme with individual timestep
\citep{Makino92} and block timestep \citep{McMillan86} for field
stars, MBHs, and the relative motion in binary MBHs.  As for the
motion of the center of mass of MBH binaries, a fourth-order Hermite
Ahmad-Cohen scheme \citep{Makino92} is employed.  In an Hermite
Ahmad-Cohen scheme, the acceleration due to single MBHs or stars near
a binary MBH, and those due to distant stars are calculated on
separate timesteps, which we call "neighbor step" and "distant step",
respectively.

The timestep except the neighbor step is determined as
\begin{equation}
  \Delta t = \sqrt{\eta f({\bf a})}, \label{eq:timestep1}
\end{equation}
where $\eta$ is the accuracy parameter, and ${\bf a}$ is the
acceleration of a field star, a single MBH, or the relative motion of
a binary MBH. The function $f$ in equation (\ref{eq:timestep1}) is
given by
\begin{equation}
  f({\bf x}) = \frac{|{\bf x}||{\bf x^{(3)}}|+|{\bf x^{(2)}}|^2}{|{\bf
      x}^{(1)}||{\bf x^{(4)}}|+|{\bf x^{(3)}}|^2}.
\end{equation}
The accuracy parameter is set to be $\eta=0.01$ for timestep of field
stars, $\eta=0.0025$ for timestep of single MBHs and distant step of
the center of mass of binary MBHs, and $\eta=0.000625$ for the
relative motion in binary MBHs. The timestep for the center of mass of
a binary MBH is determined as
\begin{equation}
  \Delta t = \sqrt{\min \left[ \eta_1 f({\bf a}), \eta_2 f({\bf
        a}_{\rm PN}) \right]},
\end{equation}
where $\eta_1=0.0025$ and $\eta_2=0.000625$ are the accuracy
parameters, and $a_{\rm PN}$ is the acceleration due to post-Newtonian
corrections for the center of mass of a binary MBH.  The relative
error in our simulations is less than $0.1$ \% regarding the total
energy.

We perform simulations with $64$ nodes of the FIRST simulator in
University of Tsukuba \citep{Umemura08}.  At each node, the FIRST
simulator is equipped with one Blade-GRAPE board, which is the
accelerator of the gravity calculations for collisional $N$-body
problem.  The gravity by field stars for a given field star and MBH is
calculated in parallel.

\section{Results}
\label{sec:rslt}

\subsection{Merger of multiple MBHs}

We calculate a system of ten MBHs in one galaxy during about $140$
$N$-body time units, which corresponds to about 800 Myr in physical
units. We find that several MBHs merge into one for all simulations.
We summarize MBH masses at the final time of simulations in Table
\ref{tab:sum}. The heaviest MBH has $m_{\rm B,max} = 4$ - $6 \times
10^{-4}$, while the second heaviest MBH has $m_{\rm B,sec} = 1 \times
10^{-4}$, except model A$_1$. The growth of such a dominant MBH is
weakly dependent on the initial stellar mass density around MBHs.

\begin{table}
  \caption{Summary for our simulation results.}
  \begin{center}
    \begin{tabular}{c|cc|cc}
      \hline
      Model name & $r_{\rm MBH} / r_{\rm v}$ & Rand \# & $10^4 m_{\rm B,max}$ & $10^4 m_{\rm B,sec}$ \\
      \hline
      \hline 
      A$_1$ & $1/3$ & R$1$ & $4$ & $3$ (ejected) \\
      A$_2$ & $1/3$ & R$2$ & $4$ & $1$ \\
      A$_3$ & $1/3$ & R$3$ & $6$ & $1$ \\
      B     & $2/3$ & R$1$  & $5$ & $1$ \\
      C     & $1$   & R$1$  & $5$ & $1$ \\
      \hline
    \end{tabular}
  \end{center}
  \label{tab:sum}
\end{table}

\begin{figure}
  \begin{center}
    \includegraphics[scale=1.0]{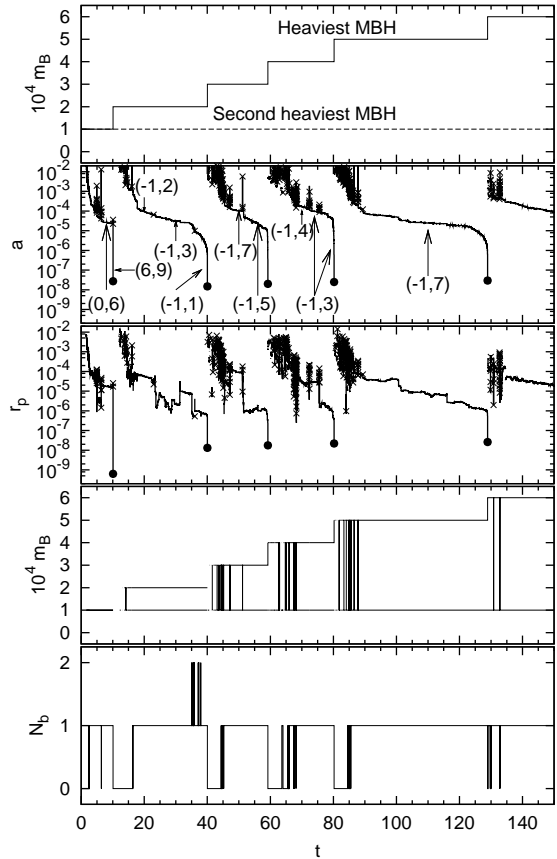}
  \end{center}
  \caption{Time evolution of MBHs in model A$_3$. From top to bottom,
    the panels show the mass of the heaviest MBH and second heaviest
    at each time, the semi-major axis, the distance at the pericenter,
    and the masses of two MBHs when a binary forms, and the number of
    binary MBHs. As for the number of binary MBHs, we only count
    binary MBHs whose semi-major axes are less than $10^{-3}$.  Pairs
    of integers in the second panel show the labels of MBHs composing
    the binary MBHs, where the heaviest MBH is labeled with "$-1$". We
    attached labels only to binary MBHs which are long-lived, or merge
    eventually.  In the second top and middle panels, filled circles
    indicate the moments when MBHs merge and crosses denote those when
    binary components are exchanged. }
  \label{fig:bbh}
\end{figure}

We see the process in which only one MBH grows in each simulation,
using the result of model A$_3$ as a typical case. As shown in the top
panel of Figure \ref{fig:bbh}, only one MBH grows, and other MBHs do
not grow.  The second and third panels of Figure \ref{fig:bbh} show
that MBHs compose a binary with a semi-major axis less than $10^{-4}$
for a longer time than the dynamical time ($t_{\rm dy}\sim 1$).  When
a binary merges to form a heavier MBH, a lighter component is often
exchanged by a third MBH.  Thereafter, a binary MBH forms again,
containing the heaviest MBH (see the fourth panel of Figure
\ref{fig:bbh}). This is because a heavier MBH is easier to be retained
in a binary MBH through an interaction between the binary MBH and
single MBH. Consequently, only one heavy MBH grows in a galaxy.

The reason why no other MBHs merge is understood as follows.  When a
binary MBH forms in a galaxy, the binary prevents the formation of
another binary MBH, because the preformed binary MBH gives other MBHs
kick velocities.  Since MBHs cannot merge without forming a binary
MBH, no other MBHs can merge.

In model A$_1$, the first binary forms just temporarily.  It is
ejected with a speed of more than $1000$ km/s due to the back reaction
of sling-shot mechanism, when the binary with masses of $2 \times
10^{-4}$ and $1 \times 10^{-4}$ merges through the sling-shot
mechanism of an interaction MBH.  After the ejection, the second
binary forms and merge to produce the MBH with mass of $4 \times
10^{-4}$ through the same process as described above.  In our
simulations, the ejection by the sling-shot interaction is observed
only once.  In all the simulations, the merger of binary MBHs occurs
$21$ times in total. Hence, the ejection of the heaviest MBH due to
the sling-shot seems rare.

\subsection{Merger mechanism of a binary MBH}
\label{sec:bmerge}

\begin{figure}
  \begin{center}
    \includegraphics[scale=1.0]{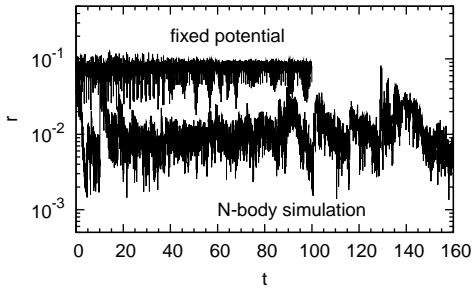}
  \end{center}
  \caption{Time variation of the distance of the third nearest MBH
    from the galactic center. The result in model A$_3$ and that in a
    fixed potential are compared.}
  \label{fig:friction}
\end{figure}

For the mergers of the MBHs, the dynamical friction plays a key
role. In Figure \ref{fig:friction}, the time variation of the distance
of the third nearest MBH from the galactic center is shown, where the
result in the $N$-body simulation and that in a fixed potential of the
Hernquist model are compared. In the fixed potential, no dynamical
friction is exerted.  As seen in this figure, the dynamical friction
by field stars in the $N$-body simulation allows the MBHs to gather
near the galaxy center. Thus, two MBHs can compose a binary MBH, and
subsequently another MBH can intrude the binary MBH. In contrast, in a
fixed potential, even a binary MBH can not be formed, since they can
not gather at around the galaxy center.

\begin{figure}
  \begin{center}
    \includegraphics[scale=1.0]{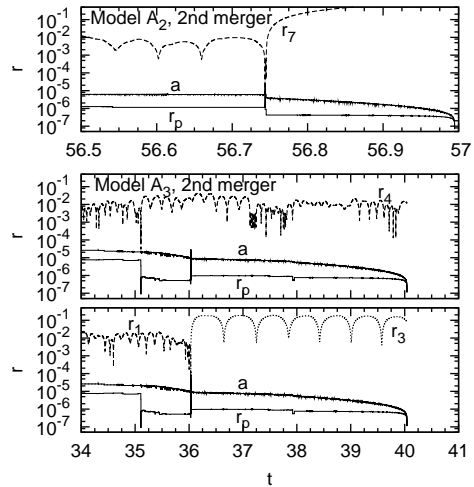}
  \end{center}
  \caption{Time variation of the semi-major axes, $a$, and the
    distance at the pericenter of binary MBHs, $r_{\rm p}$, just
    before the second mergers in models A$_2$ (top) and A$_3$ (middle
    and bottom).  Also, the separation between the center of mass of
    the binary MBHs and single MBHs which interact strongly with the
    binary MBHs is shown by dashed and dotted curves ($r_7$, $r_4$,
    $r_1$, and $r_3$), where the numbers attached to $r$ are the
    labels of MBHs.  The bottom panel demonstrates the exchange
    interaction that one binary component is replaced by another
    single MBH.}
  \label{fig:3body-gw}
\end{figure}

Here we see the details of the merger mechanism of binary MBHs. In
Figure \ref{fig:3body-gw}, the processes of the second merger of MBHs
are shown in models A$_2$ and A$_3$. The second merger in model A$_2$
is the simplest case. The binary MBH and a single MBH approaches to
each other, and strongly interact. Consequently, the distance of the
binary MBH at the pericenter shrinks, followed by emitting
gravitational wave radiation and merger. In practice, another single
MBH often intrudes into the binary MBH whose orbit is being decayed
due to energy loss through gravitational wave radiation, and
subsequently the semi-major axis and eccentricity of the binary MBH
are changed, as seen in the middle and bottom panels of figure
\ref{fig:3body-gw}. However, the crucial impact is brought by one
strong interaction. In our five simulations, there does not occur the
simultaneous interaction of multiple MBHs that triggers the merger of
a MBH binary.

Also, we have not observed the secular angular momentum loss of a
binary MBH through the Kozai mechanism \citep{Kozai62}.  The Kozai
mechanism can occur, only if the internal orbit of the binary MBH is
closed in every binary period. However, the internal orbit is not
closed due to the relativistic pericenter shift (1PN and 2PN)
\citep{Blaes02,Berentzen09}.  Actually, we have found that the Kozai
mechanism works for a binary to merge only if we do not include the
1PN and 2PN terms.  The suppression of the Kozai mechanism is also
demonstrated in the case of stellar-sized black holes \citep{Miller02}
and in the planetary orbits \citep{Fabrycky07}.

\section{Discussion}
\label{sec:dis}

The present simulations have the scalings shown in equations
(\ref{scaling1}), (\ref{scaling2}), and (\ref{scaling3}).  Thus, the
present results are applicable for the different mass scales of a host
galaxy and MBHs, if the density of the host galaxy satisfies the
scaling.  For instance, if the collapse redshift of a host galaxy is
shifted from $z=6$ to $z=10$, the virial density of the host galaxy is
expected to be increased by a factor of $3.9$. Hence, the same merger
processes of MBHs are expected for a host galaxy with $2.6\times
10^{10}M_{\odot}$ and MBHs with equal mass of $2.6\times
10^{6}M_{\odot}$.

In the present model, the host galaxy is assumed to have a spherical
Hernquist profile.  MBHs ejected from the center on a nearly radial
orbit with speed below that of escape speed repeatedly return to the
center in our spherical model. However, these MBHs do not contribute
much to the merger of binary MBHs.  We have found that there is no
time for such MBHs to interact with binary MBHs, since they pass the
center with high speed. MBHs which gather towards the galactic center
by the dynamical friction more effectively contribute to the merger of
a binary MBH.

We have ignored the recoil by gravitational wave. Unless black hole
spins are aligned, the recoil velocity can reach up to $4000$ km/s
\citep{Campanelli07,Lousto10}. Such a large recoil velocity can eject
the merger remnant from the galaxy. However, if their spins are
aligned before their merger due to relativistic spin precession
\citep{Kesden10}, then the recoil velocity decreases to a few $100$
km/s. If the recoil velocity is a few $100$km/s, the merger remnant is
possibly confined in a host galaxy with the velocity dispersion of
$300$ km/s.  Then, the recoil may not the results dramatically.
Nonetheless, it seems worth exploring carefully the effects of recoil,
which will be considered in the future analysis.

We have assumed that a binary MBH immediately merges at the moment
when their separation become smaller than ten times the sum of their
Schwarzschild radii. In practice, a binary MBHs take a bit more time
to merge. But, the timescale is too short, the merger to be disturbed
by the intrusion of another MBH. The merging timescale of a binary MBH
is $\sim 10^{-12}$ in $N$-body units, while it interacts with a MBH
once a dynamical time of the galaxy, $\sim 1$ in $N$-body
units. Actually, as seen in Figure \ref{fig:3body-gw}, any MBH does
not approach to the binary MBHs after the semi-major axis becomes
about $10^{-7}$ length unit.

\section{Summary}
\label{sec:sum}

We have performed highly-accurate $N$-body simulations to explore the
evolution of multiple MBHs in one galaxy.  Here, ten MBHs with equal
mass of $10^{7}M_{\odot}$ are set in a galaxy with
$10^{11}M_{\odot}$. As a result, it is found that $4$ - $6$ MBHs
successively merge, resulting in a single heavier MBH within 1 Gyr.
The growth timescale is shorter than a Hubble time at redshift $z \sim
6$.  After $4$ - $6$ MBHs merge, the other MBHs in the galaxy have the
same masses as those at the initial time.

The key physics for the successive merger is the dynamical friction
that allows the formation of binary MBHs and frequent interactions of
single MBHs with the binary MBHs at the galactic center.  Hence, the
key parameter for the MBH merger is the density of field stars in the
regions where MBHs are distributed. The distance of a MBH binary at
the pericenter shrinks through the sling-shot mechanism of another
MBH. It followed by the shrink of the semi-major axis due to the
energy loss by gravitational wave radiation, and eventually the binary
merges.  We have found that the secular angular momentum loss by the
Kozai mechanism does not work for a binary to merge if the
relativistic periastron shift is properly included.  The present
simulations imply that a large MBH can be formed from the system of
multiple MBHs with smaller mass purely through $N$-body process.  The
present results are applicable for the different mass scales of a host
galaxy and MBHs, if the density of the host galaxy satisfies the
scaling in the $N$-body units.  This merger path can be important for
the growth of MBHs in a primordial galaxy.  Also, the MBH merger may
be a potential source of gravitational waves for the Laser
Interferometer Space Antenna (LISA) and pulsar timing.

\acknowledgments

We thank Toshiyuki Fukushige for helpful advice on parallel $N$-body
code, Takashi Okamoto for fruitful discussion on initial conditions,
Masaki Iwasawa for useful comments about our simulation method, and
anonymous referee for meaningful advice on post-Newtonian
approximation.  Numerical simulations have been performed with
computational facilities at the Center for Computational Sciences in
the University of Tsukuba. This work was supported in part by the
FIRST project based on the Grants-in-Aid for Specially Promoted
Research by MEXT (16002003), and Grant-in-Aid for Scientific Research
(S) by JSPS (20224002).

\end{document}